\documentclass[12pt]{article}
\usepackage[dvips]{graphicx}
\usepackage{amsmath,amssymb,amsfonts}

\begin{document}

\title{Regeneration of $K^0_{S}$ mesons}
\author{V.I. Nazaruk\\
Institute for Nuclear Research of RAS, 60th October\\
Anniversary Prospect 7a, 117312 Moscow, Russia.*}

\date{}
\maketitle
\bigskip

\begin{abstract}
It is shown that in the previous calculations of $K^0_{S}$ regeneration the noncoupled equations of motion have been considered instead of coupled one. We present the calculations based on the exact solution of coupled equations of motion and perturbation theory. The results differ radically from the previous ones.

\end{abstract}

\vspace{5mm}

\vspace{5mm}

\vspace{1cm}

*E-mail: nazaruk@inr.ru

\newpage
\setcounter{equation}{0}
\section{Introduction}
The effect of kaon regeneration is known since the 1950s. The motive of our paper is as follows. In the previous calculations [1-7] a system of non-coupled equations of motion has been considered instead of the coupled ones (see below). This is a fundamental defect since it leads to a qualitative disagreement in the results [8,9]. This means that regeneration has been not described at all. The main goal of this paper is to study the connection between approaches based on exact solution [8] and perturbation theory [9].

Let $K^0_{L}$ fall onto the plate at $t=0$. We use notations of Ref. [5]. Since 
\begin{equation}
\left.\mid\!K^0_L\right>=\left.(\mid\!K^0\right>+\left.\mid\!\bar{K}^0\right>)/\sqrt{2},
\end{equation}
the evolution of $K^0_{L}$ in the medium is described by the following equation:
\begin{equation}
\left.\mid\!K'_L(t)\right>=(\left.\mid\!K^0(t)\right>+\left.\mid\!\bar{K}^0(t)\right>)/\sqrt{2}=[\left.\mid\!K^0\right>K^0(t)+\left.\mid\!\bar{K}^0\right>\bar{K}^0(t)]/\sqrt{2}.
\end{equation}
Here $\left.\mid\!K^0\right>$ and $\left.\mid\!\bar{K}^0\right>$ are the states of $K^0$ and $\bar{K}^0$, respectively; $K^0(t)$ and $\bar{K}^0(t)$ are the amplitudes of states (spatial wave functions) of  $K^0$ and $\bar{K}^0$, respectively. $K^0(t)$ and $\bar{K}^0(t)$ are calculated in $K^0,\bar{K}^0$ representation.

With these $K^0(t)$ and $\bar{K}^0(t)$ we revert to $K^0_{L},K^0_{S}$ representation:
\begin{eqnarray}
\left.\mid\!K'_L(t)\right>=\frac{1}{\sqrt{2}}[(\left.\mid\!K_L\right>+\left.\mid\!K_S\right>)K^0(t))/\sqrt{2}+
(\left.\mid\!K_L\right>-\left.\mid\!K_S\right>)\bar{K}^0(t))/\sqrt{2}]=\nonumber\\
\left.K_L(t)\mid\!K_L\right>+\left.K_S(t)\mid\!K_S\right>,  
\end{eqnarray}
where
\begin{eqnarray}
K_L(t)= \frac{1}{2}[K^0(t)+\bar{K}^0(t)],\nonumber\\
K_S(t)= \frac{1}{2}[K^0(t)-\bar{K}^0(t)].
\end{eqnarray}
$\mid \!K_S(t)\!\mid ^2$ is the probability of finding $K^0_{S}$. 

Let us calculate $K^0(t)$ and $\bar{K}^0(t)$. The coupled equations for zero momentum $K^0$ and $\bar{K}^0$ in the medium are the following:
\begin{eqnarray}
(i\partial_t-M)K^0=\epsilon \bar{K}^0,\nonumber\\
(i\partial_t-(M+V))\bar{K}^0=\epsilon K^0,
\end{eqnarray}
where
\begin{eqnarray}
M=m_{K^0}+U_{K^0}-i\Gamma _{K^0}^d/2,\nonumber\\
V=(m_{\bar{K}^0}-m_{K^0})+(U_{\bar{K}^0}-U_{K^0})-(i\Gamma _{\bar{K}^0}^d/2-i\Gamma _{K^0}^d/2).
\end{eqnarray}
Here $\epsilon =(m_L-m_S)/2=\Delta m/2$ is a small parameter, $U_{K^0}$ and $U_{\bar{K}^0}$ are the potentials of $K^0$ and ${\bar{K}^0}$, $\Gamma _{K^0}^d$ and $\Gamma _{\bar{K}^0}^d$ are the decay widths of $K^0$ and ${\bar{K}^0}$, respectively.

Equations (5) follow uniquely from the unperturbed and interaction Hamiltonians:
\begin{eqnarray}
H_0=-\nabla^2/2m+ U_{K^0},\nonumber\\
H_I=H_{K^0\bar{K}^0}+H_W+V,\nonumber\\
H_{K^0\bar{K}^0}=\int d^3x(\epsilon \bar{\Psi }_{\bar{K}^0}(x)\Psi _{K^0}(x)+H.c.).
\end{eqnarray}
Here $H_{K^0\bar{K}^0}$ and $H_W$ are the Hamiltonians of the $K^0\bar{K}^0$
conversion and decay of the $K$-mesons, respectively; $\bar{\Psi }_{\bar{K}^0}$ and $\Psi _{K^0}$ are the fields of $\bar{K}^0$ and $K^0$, respectively. 

Equations (5) are coupled ones due to off-diagonal mass $\epsilon $. In the previous old calculations [2] the starting equations are (see Eqs. (3) from [2]):
\begin{eqnarray}
(\partial_x-ink)K^0=0,\nonumber\\
(\partial_x-in'k)\bar{K}^0=0,
\end{eqnarray}
where $n$ and $n'$ are the indexes of refraction for $K^0$ and $\bar{K}^0$, 
respectively. In notations of Ref. [2] $K^0=\alpha $ and $\bar{K}^0=\alpha '$,
$K^0_{S}=\alpha _1$ and $K^0_{L}=\alpha _2$. In above-given Eq. (8) we substitute $K^0=(\alpha _1+ i\alpha _2)/ \sqrt{2}$, $\bar{K}^0=(\alpha _1-i\alpha _2)/ \sqrt{2}$ and include the effect of the weak interactions as in [2]. We obtain Eq. (5) and result (6) from Ref. [2].

So the starting equations (3) in [2] are noncoupled. There is no off-diagonal mass $\epsilon =(m_L-m_S)/2$. This is a fundamental defect. The noncoupled equations exist only for the stationary states and don't exist for $K^0$ and $\bar{K}^0$. For $n\bar{n}$ transitions in the medium [10-15] the coupled equations are solved as well as for any $ab$-oscillations.

The value of $\mid \!K_S(t)\!\mid ^2$ is of particular interest. We find 
$K^0(t)$ and $\bar{K}^0(t)$ and exact expression for $\mid \!K_S(t)\!\mid ^2$
[8]. Our main concern here is comparison with calculation based on perturbation theory. For this purpose we consider the particular case of exact solution. We put $\Gamma _{K^0}^d= \Gamma _{\bar{K}^0}^d=\Gamma ^d$, $m_{K^0}=m_{\bar{K}^0}=m$ ($m$ is the mass of $K^0$) and denote:
\begin{equation}
\Delta \Gamma =\Gamma _{\bar{K}^0}^a-\Gamma _{K^0}^a,
\end{equation}
where $\Gamma _{K^0}^a$ and $\Gamma _{\bar{K}^0}^a$ are the widths of absorption (not decay) of $K^0$ and ${\bar{K}^0}$, respectively.

If $\Delta \Gamma t\gg 1$, the exact expression for $\mid \!K_S(t)\!\mid ^2$ [8] has the form
\begin{eqnarray}
\mid \!K_S(t)\!\mid ^2\approx \frac{1}{4}e^{-\Gamma (K_L\rightarrow K_S)t}e^{-(\Gamma ^a_{K^0}+\Gamma ^d)t},\nonumber\\
\Gamma (K_L\rightarrow K_S)=\frac{\epsilon ^2}{\mid V\mid ^2}\Delta \Gamma,
\end{eqnarray}
where $\Gamma (K_L\rightarrow K_S)$ is the width of $K^0_{L}K^0_{S}$ transition. 
It is significant that ${\rm Re}V\ne 0$ in contrast to [8]. 

The calculation presented above is cumbersome and formal. The verification is needed. In [9] the approach based on perturbation theory has been proposed. The regeneration followed by decay $K^0_L\rightarrow K^0_S\rightarrow \pi \pi $ is considered. From (5) the process amplitude $M(K^0_L\rightarrow K^0_S\rightarrow \pi \pi )$ is found to be
\begin{equation}
M(K^0_L\rightarrow K^0_S\rightarrow \pi \pi )=
\frac{\epsilon }{V}M_d(K^0_S\rightarrow \pi \pi).
\end{equation}
Here $M_d(K^0_S\rightarrow \pi \pi)$ is the in-medium amplitude of the decay $K^0_S\rightarrow \pi \pi$. The corresponding process width is
\begin{equation}
\Gamma (K^0_L\rightarrow K^0_S\rightarrow \pi \pi)=\frac{\epsilon ^2}{\mid V\mid ^2}\Gamma _d(K^0_S\rightarrow \pi \pi ),
\end{equation}
where $\Gamma _d(K^0_S\rightarrow \pi \pi )$ is the width of decay $K^0_S\rightarrow \pi \pi $.

Consider now the connection between the models based on diagram technique and exact solution. To do this we write (12) in the form
\begin{equation}
\Gamma (K^0_L\rightarrow K^0_S\rightarrow \pi \pi)=\frac{\epsilon ^2}{\mid V\mid ^2}\Gamma _d(K^0_S\rightarrow \pi \pi )\frac{\Delta \Gamma }{\Delta \Gamma }=
\Gamma (K_L\rightarrow K_S)W,
\end{equation}
\begin{equation}
W=\frac{\Gamma _d(K^0_S\rightarrow \pi \pi )}{\Delta \Gamma },
\end{equation}
where $W$ is the probability of the $K^0_S$ decay on the channel 
$K^0_S\rightarrow \pi \pi $. The physical sense of (13) is obvious: the multistep process $K^0_L\rightarrow K^0_S\rightarrow \pi \pi $ involves the subprocess of $K_LK_S$ transition (regeneration). Equation (13) is verification of the approaches given above.

Finally, both of calculations give the same result for the case $\Delta \Gamma t\gg 1$. The result obtained by means of exact solution should be studied first since it is valid for any value of $\Delta \Gamma t$. We will continue our consideration in the following paper.

\end{document}